\pdfoutput=1
\documentclass[12pt,a4paper]{article}      

\usepackage[bookmarks=false]{hyperref}
\usepackage{amsthm,amsmath,amssymb,xcolor,sectsty,latexsym,mathtools,mathrsfs}
\usepackage{slashed}  
\usepackage{bm}
\usepackage{epsfig}
\usepackage{dcolumn}
\definecolor{vermelho}{cmyk}{0,.88,.77,.40}
\usepackage{cite}
\usepackage{feynmp}
\numberwithin{equation}{section}
\usepackage[textwidth=6.3in,top=1.3in,bottom=1.3in]{geometry}
\usepackage{setspace}
\chapterfont{\color{vermelho}}
\usepackage{simplewick}

\newcommand{\be}{\begin{equation}}
\newcommand{\ee}{\end{equation}}
\newcommand{\beq}{\begin{equation}}
\newcommand{\eeq}{\end{equation}}
\newcommand{\ba}{\begin{eqnarray}}
\newcommand{\ea}{\end{eqnarray}}
\newcommand{\bef}{\begin{figure}}
\newcommand{\eef}{\end{figure}}

\newcommand{\p}{\partial}
\newcommand{\al}{\alpha}
\newcommand{\ep}{\epsilon}

\newcommand{\si}{\sigma}

\newcommand{\g}{\gamma}

\newcommand{\cL}{{\cal L}}

\newcommand{\nn}{\nonumber}

\begin{document}

\thispagestyle{empty}
\begin{titlepage}
\nopagebreak

\title{  \begin{center}\bf Patient Observers and Non-perturbative Infrared Dynamics in Inflation \end{center} }

\vfill
\author{Ricardo Z. Ferreira$^{a}$\footnote{rferreira@icc.ub.edu}, ~ McCullen Sandora$^{b}$\footnote{mccullen.sandora@tufts.edu}~ and ~  Martin S. Sloth$^{c}$\footnote{sloth@cp3.sdu.dk}
}
\date{ }

\maketitle

\begin{center}
	\vspace{-0.7cm}
	{\it  $^a$Departament de Fisica Fonamental i Institut de Ciencies del Cosmos}\\
	{\it  Universitat de Barcelona, Marti i Franques, 1, 08028, Barcelona, Spain}\\
	\vspace{0.2cm}
	{\it  $^b$Institute of Cosmology, Department of Physics and Astronomy}\\
	{\it  Tufts University, Medford, MA 02155, USA}\\
	\vspace{0.2cm}
	{\it  $^c$CP$^3$-Origins, Center for Cosmology and Particle Physics Phenomenology}\\
	{\it  University of Southern Denmark, Campusvej 55, 5230 Odense M, Denmark}\\
	
\end{center}
\vfill
\begin{abstract}
We have previously derived the effect of soft graviton modes on the quantum state of de Sitter using spontaneously broken asymptotic symmetries. In the present paper we reinterpret this effect in terms of particle production and relate the quantum states with and without soft modes by means of Bogoliubov transformations. This also enables us to address the much discussed issues regarding the observability of infrared effects in de Sitter from a new perspective. While it is commonly agreed that infrared effects are not visible to a single sub-horizon observer at late times, we argue that the question is less trivial for a {\it patient observer} who has lived long enough to have a record of the state before the soft mode was created. Though classically there is no obstruction to measuring this effect locally, we give several indications that quantum mechanical uncertainties may censor the effect.  We then apply our methods to find a non-perturbative description of the quantum state pertaining to the Page time of de Sitter, and derive with these new methods the probability distribution for the local quantum states of de Sitter and slow-roll inflation in the presence of long modes. Finally, we use this to formulate a precise criterion for the existence of eternal inflation in general classes of slow-roll inflation.
\end{abstract}
\noindent
DNRF90
\hfill \\
\vfill
\end{titlepage}

\section{Introduction}
Though inflation is commonly invoked as the mechanism in the early universe that solves the horizon and flatness problems by exponentially diluting away any inhomogeneities, it has been argued that if it runs for long enough it actually does not produce such a flat, smooth spacetime.  This is because it also generates quantum mechanical fluctuations that get stretched to superhorizon lengths, and if these accrue the spacetime gets warped on very long length scales.  The time it takes for this to occur is the Page time of de Sitter\footnote{We define the Page time of de Sitter as $t_{dS}\sim R_{dS}S_{dS}\sim M_p^2/H^3$, where $R_{dS}$ is the de Sitter radius and $S_{dS}$ is the de Sitter entropy.} in analogy with the black hole case, as has long been argued.  Recently we showed that these changes can be encapsulated by a change in the quantum state of the system for the de Sitter case, and showed that after the Page time the perturbative description in the initial vacuum state breaks down, forcing the overlap between the initial and final states to go to zero \cite{1609.06318}. This elaborates on the findings of \cite{1005.1056,Giddings:2011ze,Giddings:2011zd} who, for example in  \cite{1005.1056}, demonstrated that the circumsphere of de Sitter compactified on a torus undergoes a large variance due to infrared fluctuations after a similar time scale, and of \cite{Dvali:2013eja,Dvali:2014gua,Dvali:2017eba}\footnote{This time-scale also appeared in the discussion of the thermalization time scale of de Sitter in \cite{Danielsson:2003wb}} who also identified this time scale in de Sitter as a point where the perturbative treatment appears to break down.

\begin{centering}
	\begin{figure*}[h]
		\centering
		\includegraphics[width=6cm]{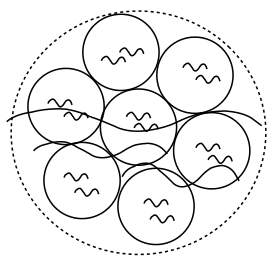}
		\caption{The change in geometry from the emission of a long wavelength mode at late times.  A local observer only has access to information contained within their horizon, and the effect of long modes can be gauged away locally.  However, the difference between different causally separated local observers becomes large. To observe this effect, inflation needs to come to an end, and long wavelength modes need to re-enter the horizon.}
		\label{pepsi}
	\end{figure*}
\end{centering}

\begin{centering}
	\begin{figure*}[h]
		\centering
		\includegraphics[width=6cm]{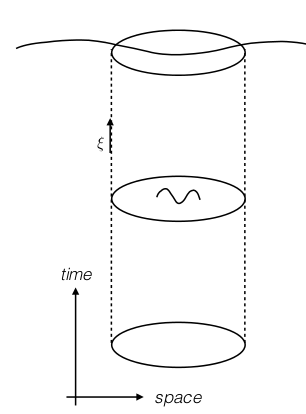}
		\caption{The change in geometry from the emission of a long wavelength mode.  A local observer only has access to information contained within their horizon, and the effect of long modes can be gauged away locally.  However, the difference between initial and final gauges can be kept track of, in which case even a physical observer would be able to measure the effect once it becomes large.}
		\label{can}
	\end{figure*}
\end{centering}

At first this seems at odds with the fact that a physical observer still experiences a spacetime that locally resembles a flat, homogeneous and isotropic spacetime.  Indeed, because this breakdown is a property of the global description of the state, it has been argued that the effect is unphysical. Indeed, any mode with wavelength much larger than the system being observed can be locally gauged away, as a consequence of the equivalence principle, and at late times super-horizon modes cannot be measured within a single Hubble patch as illustrated in Fig. \ref{pepsi}.  Any observation of this effect would require a nonlocal measurement in the sense that it would need a comparison of the effect at (minimum) two very separate spacetime points.  A single local observer is able to register the effect if they were willing to wait for long enough.  After a Page time, it would always be possible to construct a coordinate system corresponding to flat space; however, this would not coincide with the initial coordinate system.  Thus, while our observer remains in flat space the entire time, the flat spatial slices are related to each other in a nontrivial way (a diffeomorphism at asymptotic future null infinity) as we illustrate in Fig. \ref{can}.

To make sure this represents a physical effect, we can imagine constructing a device that would be able to register this change.  What we have in mind is a network of satellites, loosely tethered together to counteract the effect of expansion, initially arranged in a sphere.  After a Page time, the initial sphere would be deformed into an ellipsoid of different size by an $\mathcal{O}(1)$ amount.  This represents the memory effect associated with the asymptotic symmetry group transformations that have taken place, as explained in \cite{1411.5745}.  We could even imagine observing this effect with a single telescope, if we were able to register the stresses induced by the long wavelength modes throughout its body.

When a measuring device is created, it is done so in vacuum, and so automatically gauges away every long mode that was emitted prior to that event. Therefore, only long modes emitted after that point will leave any effect on the system. This leads to our definition of a patient observer: 
\emph{A patient observer is an observer that has been in existence prior to the time when a given long mode of interest was created, and is equipped with some measurement device, necessarily not de Sitter invariant, to record its state for the entire duration.} 
The usual curvature perturbations of wavelength shorter than the soft mode are not patient observers, since their freeze-out dynamics, that is, the time at which they cross the horizon, is set by the background geometry that includes all soft modes that have been previously emitted. This is the reason why IR effects can be gauged away in the case of the usual spectrum of curvature perturbations in cosmology, and prevents an accurate record of the history before its creation from being recorded. However, the arrangement of satellites mentioned above, as well as any isocurvature modes present, are examples of a patient observer. In section \ref{patient} we will discuss a third type of patient observer based off an Unruh detector.  In section \ref{impatience} we discuss possible fundamental limitations to constructing patient observers in practice due to quantum mechanical effects.

The breaking of de Sitter invariance by the observer plays a crucial role, and without it, a patient observer would not be able to distinguish the state before and after the soft mode is created. This is similar to the analogous example of electromagnetic memory in the case of exploding charges \cite{Susskind:2015hpa}, where the change in the state is a pure gauge transformation in the absence of the observers (in that case the superconducting nodes). However, the presence of the superconducting nodes on a sphere surrounding the initial charges spontaneously break the gauge invariance in the global state, and therefore they are able to record a memory of the state before the explosion of the charges, and register the difference afterwards. The superconducting nodes are the electromagnetic analog of a patient observer.

Though the departure from the initial vacuum becomes large at the Page time, giving an indication that the perturbative description breaks down on this timescale, the prognosis for describing the physics is not necessarily bleak if we can find a non-perturbative way of organizing these effects.  This is the procedure we outline in this paper, where we show that it is possible to explicitly evaluate the effects of long modes on correlators to all orders. While related methods have been discussed in the literature \cite{1005.1056,Giddings:2011zd,Tanaka:2013caa,Urakawa:2010it,Tanaka:2012wi,Byrnes:2010yc,Gerstenlauer:2011ti, Frob:2013ht, Burgess:2015ajz}, our technique enables us to derive known results in a compact fashion, and also find new applications. As an example, which also serves as a non-trivial check of our methods, we re-derive the probability distribution of the comoving curvature perturbation, and show that our results matches the results of \cite{1103.5876}, where it was derived using instead the Kramers-Moyal equation of the cosmological comoving curvature perturbation. Our initial expressions are within several approximation schemes which prevent them from being exact: namely, we work in the constant tilt approximation, and ignore interactions among the long modes themselves, focusing on their additive influence on short modes.  This allows us to arrive at simple expressions for the correlators we consider, but we are actually able to step beyond these approximations and extract some quantitative features in the general case.  Our main application of this technology is to write down the probability distribution of an observer measuring a given value of the power spectrum, even deep in the non-perturbative regime.  We find a power law form, with the exponent dependent on the precise nature of inflation.

To achieve this, we first recast the change in state of the system as a Bogoliubov transformation in section 2.  This is shown to induce changes in the mode functions of the short wavelength modes, equivalent to the change in state obtained from the Noether charge we found in \cite{1609.06318}.  This state is then used to compute the expected values of several common correlators in section 3, and the full distribution in section 4.  We find that the averages are not very representative of a typical observation, a consequence of the system's power law behavior.

Before we begin, we comment on the concept of the Page time in inflationary spacetimes, as the generalization from de Sitter space contains some subtlety.  The most primitive definition we will need is that the number of degrees of freedom emitted rivals the degrees of freedom of the horizon, which for pure de Sitter corresponds to a number of e-folds $N\sim M_p^2/H^2$.  In slow roll spacetimes, however, the horizon size increases as well, and generically at a faster rate than long wavelength modes are emitted, so the Page time is never reached.  The minimum requirement for the Page time to make sense is that the number of horizon degrees of freedom increases at a slower rate than the emission rate.  Then, using the slow roll equation $dH/dN=-\epsilon H$, we find that $d A/dN= \epsilon A$, and, using $P_\gamma=H^2$, $P_\zeta=P_\gamma/(16\epsilon)$, we find that the change in horizon area (and therefore the number of holographic degrees of freedom)  in one e-fold is equal to $\Delta A=1/P_\zeta$.  Comparing this to the number of gravitons emitted per e-fold, $\Delta N_\gamma=2$, we see that the Page time is only possibly reached in the $P_\zeta\sim1$ regime, i.e. eternal inflation.  Note that this is only a necessary condition for the Page time to be reached:  this regime must still persist for long enough that the emitted modes actually overtake the horizon's.  During inflation, however, more scalar degrees of freedom are emitted than tensors, and so a more appropriate comparison would be to $\Delta N_\zeta=1/\epsilon$.  This yields the condition $r=P_\zeta$, which, as we will see, exactly corresponds to the condition that the loop corrections to the graviton two point function are of the same order of magnitude as the tree level contribution, or, in other words, the onset of the non-perturbative regime.

\section{Bogoliubov}\label{bogo}

In this section we show how a soft graviton or inflaton mode can be reinterpreted as inducing particle production through a Bogoliubov transformation.  We demonstrate the utility of this framework by easily computing one loop corrections to the scalar two point function, and verify that it is equivalent to a change in the state of the system, proving its validity.

Generically, we will be interested in correlators between the curvature perturbation $\zeta$ and gravitons $\gamma$ of the form
\beq
\left\langle \prod_{i=1}^n\zeta_{\tilde{k}_i}\prod_{j=1}^m\gamma_{\tilde{k}_j}\right\rangle \, .
\eeq
We choose to work in the uniform density gauge, where the effect of the long mode is to shift the momentum $k^2 \rightarrow\tilde{k}^2=e^{2\zeta_L}\left[e^{\gamma_L}\right]_{ij} k_i k_j$.  Intuitively, this will induce particle creation by altering the mode equation of the fluctuations.  For any scalar field, for instance, the wave function will obey the shifted wave equation
\beq
\tilde\Delta\phi=\left(a^{-3}\partial_t a^3 \partial_t -\frac{1}{a^2}e^{-2\zeta_L}e^{-\gamma_L}{}_{ij}\partial_i\partial_j\right)\phi=m^2\phi \, .
\eeq
We mostly focus on the case where $\phi$ is the canonically normalized field corresponding to $\zeta$, that is, $\zeta=(H/\dot{\bar\phi})\phi$.  From here, the long wavelength modes do not affect the separability properties of this wave operator, so that time and space can be analyzed individually (once it is frozen out).  Additional terms involving the long $\zeta$ mode would muddle the two by adding a nonzero shift vector, but these are slow roll suppressed. Furthermore, the part of this equation that depends on time derivatives is not altered at all, which means that if we expand in spatial plane waves, the mode function will depend on these through the effective wave number described above. If there is any spatial dependence in the long modes then this procedure is approximate, as plane waves will not be exact eigenfunctions of the Laplacian, but in the limit that the wavelength of the long modes is much larger than scales we are interested in it can be treated as a constant (anisotropic) rescaling of the coordinates.

\subsection{Inflation} \label{Bogoliubov transformation}

In this setting the mode equations are Hankel functions, which, when written in terms of conformal time $\eta=\int dt/a$ are
\beq
u_k(\eta)=c_{\nu} H (-\eta)^{3/2} H^+_\nu(-k\eta)\, ,
\eeq
where $c_{\nu}=\frac{\sqrt{\pi}}{2} e^{i \frac{\pi}{2}(\nu + 1/2)}$,  chosen to asymptote to positive frequency modes at early times.  The coefficient $\nu$ parameterizes the departure from exact de Sitter, which in terms of the scalar power spectrum tilt is $\nu=3/2-(n_s-1)/2$. 

The insight we draw is that in a shifted background the same mode function can be used except with the replacement of $k\rightarrow\tilde{k}$\cite{Maldacena:2002vr,Seery:2008ax,1005.1056,Giddings:2011ze}.  Then the Bogoliubov coefficients can be computed in the standard way.  The additional subtlety that normally does not occur is that these coefficients now depend on the long modes, that is, the Bogoliubov coefficients are now \emph{field dependent}.  However, if we restrict our attention to modes with wavelengths much shorter than the long mode, the operators commute and we are able to sidestep this subtlety.  The field in the shifted background is related to $U$ through 
\ba \label{alpha beta def}
\zeta_{\tilde{k}}|0\rangle=\left(\alpha_k u_k^*+\beta_k u_k\right)a_k^\dagger|0\rangle\, .
\ea
If there are multiple modes in the desired correlator, this procedure can be used for each independently.  Thus, for instance, the correction to the two point function is
\beq\label{optimus}
\langle \zeta_{\tilde{k}_1}\zeta_{\tilde{k}_2}\rangle=\langle \left|\alpha_k u_k^*+\beta_k u_k\right|^2 J \rangle  \delta^{(3)}(k_1+k_2) \, .
\eeq
The coefficient $J$ is a Jacobian factor, obtained from $\delta^{(3)}(\tilde k)=J\delta^{(3)}(k)$.  It remains to compute the quantity $\langle \left|\alpha_k u_k^*+\beta_k u_k\right|^2 J\rangle$, which is where the interesting physics comes in.  From the standard formulas we find that
\beq\label{grimlock}
\langle \left|\alpha_k u_k^*+\beta_k u_k\right|^2 J\rangle = \left\langle e^{(3-2\nu)\zeta_L}\left(e^{\gamma_L}{}_{ij}\hat{k}_i\hat{k}_j\right)^{-\nu}\right\rangle P_k\, .
\eeq
Where $\hat{k}$ is the unit vector pointed in the direction of the wavenumber at which the scalar correlator is being evaluated, and $P_k=|u_k|^2$ is the power spectrum.  This quantity is somewhat tricky, as it involves nontrivial index structure and, when the matrix exponential is expanded, depends on the long graviton in a complicated way.  To begin our analysis we reproduce the one loop results, before going on to a more sophisticated analysis.  

\subsection{One loop}

To reproduce the one loop results from the literature \cite{1005.1056} we need to expand the quantity (\ref{grimlock}) to second order in the long curvature and graviton modes.  This will automatically incorporate the two diagrams of the form drawn below.

\begin{centering}
	\begin{figure*}[h]
		\centering
		\includegraphics[width=12cm]{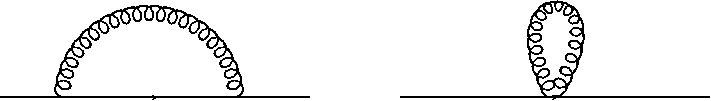}
		\caption{The two 1 loop diagrams.  The straight line is the scalar mode, and the slinky lines are both long wavelength gravitons and scalars.}
		\label{alphadecays}
	\end{figure*}
\end{centering}

To do this, let us denote $\delta_m=\gamma^m_{ij}\hat{k}_i\hat{k}_j$.  Then we have
\beq
\left\langle\hat{\tilde{k}}^{-2\nu}\right\rangle=1+\frac{(3-2\nu)^2}{2}\langle\zeta_L^2\rangle-\frac{\nu}{2}\langle\delta_2\rangle+\frac{\nu(\nu+1)}{2}\langle\delta_1^2\rangle+\mathcal{O} \left(\g_L^4, \zeta_L^4, \zeta_L^2 \gamma_L^2 \right)\, .
\eeq
To proceed we need to relate the two second order correlators to the power spectrum of gravitons. Using
\beq
\langle\gamma(p_1)_{ij}\gamma(p_2)_{kl}\rangle=P_\gamma(p_1)P_{ijkl}(p_1)\delta^{(3)}(p_1+p_2),
\eeq
where
\beq
P_{ijkl}(p)=\hat\delta_{ik}\hat\delta_{jl}+\hat\delta_{il}\hat\delta_{jk}-\hat\delta_{ij}\hat\delta_{kl},\quad \hat\delta_{ij}=\delta_{ij}-\hat{p}_i\hat{p}_j
\eeq
we find that 
\beq
\langle\delta_1^2\rangle=s^4\langle\gamma_L^2\rangle,\quad\langle\delta_2\rangle=2s^2\langle\gamma_L^2\rangle
\eeq
where $s^2=1-(\hat k\cdot\hat p)^2$.  Once we do the angular average we arrive at
\beq
\langle\delta_1^2\rangle=\frac{8}{15}\langle\gamma_L^2\rangle,\quad\langle\delta_2\rangle=\frac{4}{3}\langle\gamma_L^2\rangle
\eeq
which gives
\beq
\left\langle\hat{\tilde{k}}^{-2\nu}\right\rangle=1+\frac{(1-n_s)^2}{2}\langle\zeta_L^2\rangle+\frac{(4-n_s)(1-n_s)}{15}\langle\gamma_L^2\rangle+\mathcal{O} \left(\g_L^4, \zeta_L^4, \zeta_L^2 \gamma_L^2 \right) \, .
\eeq
We have used the fact that to lowest order $\nu=(4-n_s)/2$, which gives perfect agreement with \cite{1005.1056}.

\subsection{Charge transformation $\Leftrightarrow$ Bogoliubov transformation}

\begin{centering}
	\begin{figure*}[h]
		\centering
		\includegraphics[width=8cm]{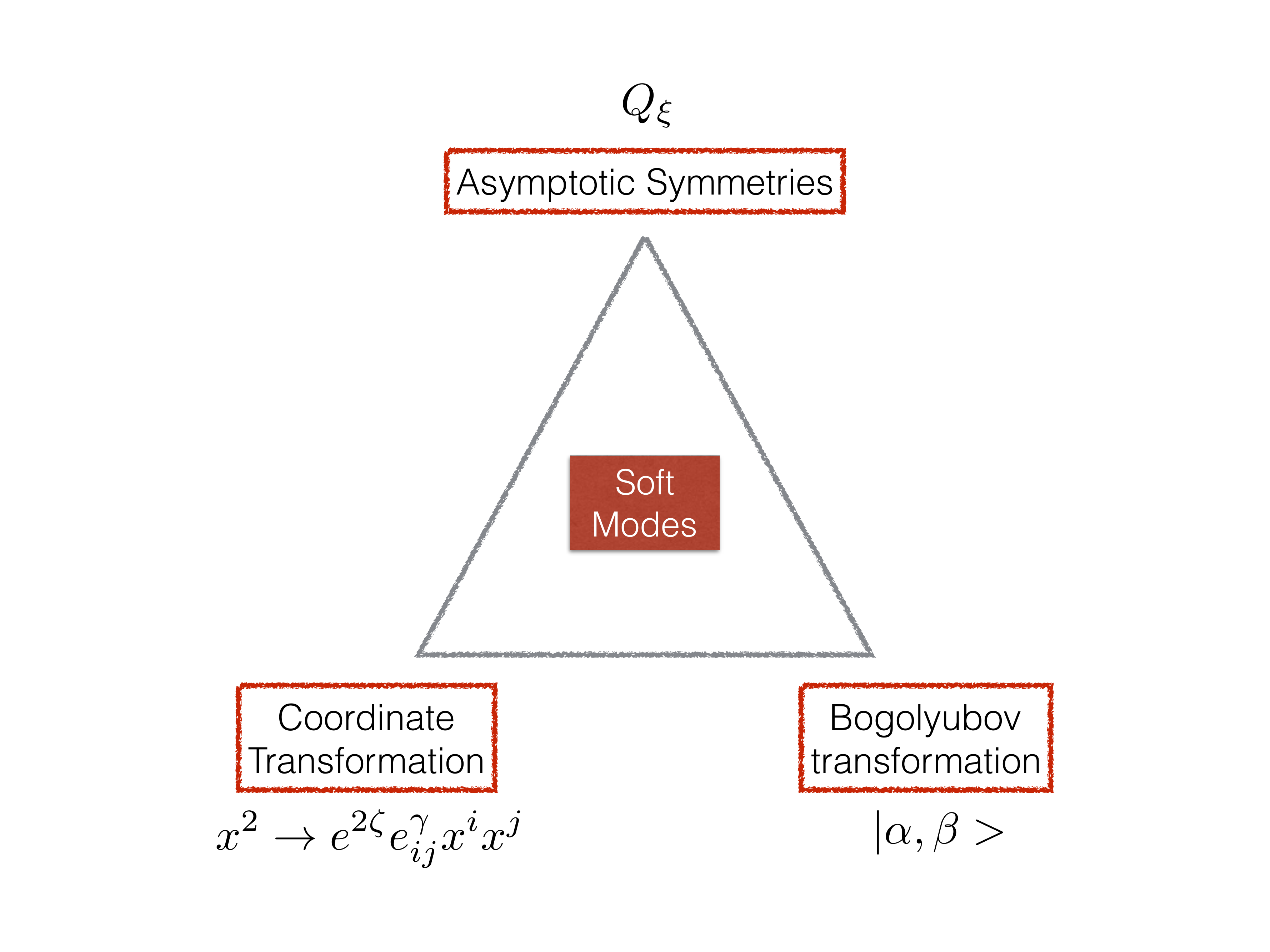}
		\caption{Equivalence between the charge associated with the asymptotic symmetries, a coordinate transformation, and a Bogoliubov transformation.}
		\label{triangle}
	\end{figure*}
\end{centering}

To close the consistency triangle of figure \ref{triangle} we need to show that the charge associated with the soft mode derived in \cite{1609.06318} indeed corresponds to a Bogoliubov transformation with the coefficients derived in section \ref{Bogoliubov transformation}.

As described in \cite{1609.06318} (and in several other works before \cite{1203.6351,1304.5527}, as well as in \cite{Kehagias:2017rpe} from the perspective of a dS/CFT correspondence) the symmetry of the action under large gauge transformations has an associated charge
\begin{eqnarray}
Q_\xi=\frac{1}{2} \int d^3 x \left[ \{\Pi_\zeta, \delta \zeta\}  + \{\Pi_\gamma^{ij}, \delta \gamma_{ij} \}  \right]
\end{eqnarray}
where $\Pi_{\zeta,g} \equiv \delta \cL/ \delta (\dot{\zeta}, \dot{g}_{ij}) $ are the conjugate momentum associated with the two gravitational degrees of freedom  $\zeta$ and $\gamma_{ij}$, and $\delta \zeta, \delta \gamma_{ij}$ are their transformations under a large gauge transformation. For simplicity we will only consider the charge associated with a soft graviton. 

We would like to verify that the charge acts on the vacuum as a Bogoliubov transformation which has the form
\begin{eqnarray} \label{BogTrans}
\left| 0' \right>= \prod_{k} \frac{e^{\frac{-\beta^*}{2\alpha}a^\dagger_k a^\dagger_{-k}}}{|\alpha_k|^{1/2}}  \left| 0 \right>,
\end{eqnarray}
where $\alpha$ and $\beta$ relate the creation and annihilation operators between the two different vacua as described in eq. \ref{alpha beta def}. 

We focus on the charge induced by a soft tensor, i.e. we consider a diffeomorphism of the form $\xi_i = \left(e^{\g^L/2}-\delta \right)_{ij} x_j$ with associated charge\footnote{We refer the reader to \cite{1609.06318} for the derivation of the cubic part of the charge.} 
\begin{eqnarray} \label{charge0}
Q= \frac{a^3 M_p^2}{4} \int d^3x \, \dot{\g}_{ij} D_L \g_{ij}
\end{eqnarray}
where $D_L \equiv \g^L_{ab}/2 \, x_b \partial_a$ and $\g^L_{ab}$ is the soft graviton appearing in the large gauge transformation.
When expanding in Fourier space the charge can be written as
\begin{eqnarray}
\exp{iQ}= \Pi_k \exp{\left(c_+ K_+ + c_- K_- + c_3 K_3\right)}
\end{eqnarray}
where $K_+ = a^\dagger_k a^\dagger_{-k}/2$, $K_- = a_k a_{-k}/2$ and $K_3 = (a_k^\dagger a_{-k} + a_{k} a^\dagger_{-k})/4$, and all the $c$'s are momentum dependent functions associated with each $K$. Making use of the formulas in appendix 5 of \cite{barnrad} we can rewrite the exponential of the charge as
\begin{eqnarray}
\exp{iQ}= \Pi_k \exp{\left(\Gamma_+ K_+\right)} \, \exp{\left( \log (\Gamma_3) K_3\right)} \, \exp{\left(\Gamma_- K_-\right)}
\end{eqnarray}
where
\begin{eqnarray}
\Gamma_\pm &=& \frac{ 2 c_\pm \sinh \beta}{ 2\beta \cosh \beta - c_3 \sinh \beta}  \\
\Gamma_3 &=& \left( \cosh \beta - \frac{c_3}{2 \beta} \sinh \beta \right)^{-2} \\
\beta^2 &=& \frac 1 4 c_3^2 - c_+ c_- .
\end{eqnarray}
All the $c$-functions are proportional to $\gamma_L$ and so is $\beta$. Thus, expanding the $\Gamma$s for small $\beta$ gives
\begin{eqnarray}
\Gamma_\pm &=& c_\pm + {\cal O }(\gamma_L^2) \\
\Gamma_3 &=& 1 + {\cal O }(\gamma_L)
\end{eqnarray}
This shows that to leading order in $\gamma_L$
\begin{eqnarray}
\exp{iQ}  \left| 0 \right>  = \mathcal{N} \, \Pi_k \exp{\left(c_+ K_+\right)}  \left| 0 \right>. 
\end{eqnarray}
where $\mathcal{N}$ is a normalization factor.

To close the triangle of fig. \ref{triangle} we still need to show that $c_+ = - \beta^*/\alpha$ with the $\alpha$ and $\beta$ computed in sec. \ref{Bogoliubov transformation}\footnote{Note that the factor of $1/2$ is already included in the definition of $K_+$.} . To do that we come back to our original expression in eq. \ref{charge0} and integrate by parts half of the integral 
\begin{eqnarray}
Q=\frac{a^3 M_p^2}{8}  \left[\int d^3 x D_L \left( \dot{\g}_{ij}  \g_{ij} \right) - D_L \left( \dot{\g}_{ij} \right)  \g_{ij}+  \dot{\g}_{ij} D_L \g_{ij} \right].
\end{eqnarray}
The total derivative will be evaluated at the boundary. Therefore, when going to Fourier space it will correspond to an operator of a single (absolute) momenta associated. We neglect this term in what follows given that in the large volume limit its effect on the state is negligible.

Then, after Fourier transforming the previous integral and select the terms proportional to $K_+$ we get
\begin{eqnarray} \label{charge1}
\tilde{Q}&=&-\frac{a^3 M_p^2}{4}  \int d^3 k \, \sum_{\sigma} \left[   \dot{\g}_k^* D_L \g_k^*-  \g_k ^* D_L  \dot{\g}_k^* \right] a_{k,\sigma}^\dagger a_{-k,\sigma}^\dagger   \quad, 
\end{eqnarray}
where we have used the fact that $\sum_{\si, \si'} \ep^{\si}_{ij}(k) \ep^{\si'}_{ij}(-k) = 2\delta_{\sigma, \sigma'}$, being $\sigma$ the graviton polarization. The previous equation defines $c_+$ to be
\begin{eqnarray} \label{c+}
c_{+,\sigma} =  \frac{a^3 M_p^2}{2i}   \left[   \dot{\g}_k^* D_L \g_k^*-  \g_k ^* D_L  \dot{\g}_k^* \right] 
\end{eqnarray}
where the $\sigma$ index just characterizes the fact that we would have a $c_+$ for each polarization.

In order to make the comparison with the Bogoliubov coefficients derived from the coordinate transformation we use the definition
\begin{eqnarray} \label{beta}
\beta &=& \frac{u'_k v_k - v'_k u_k }{W} \\ \label{alpha}
\alpha &=& \frac{u^{'}_k v^*_k - u_k v_k^{'*} }{W}
\end{eqnarray}
where $v(k)= u(\tilde{k})$ is the mode function in the shifted coordinates with $\tilde k^2= e^{2\zeta_L} e^{\gamma_L}_{ij} k_i k_j$, and $W$ is the Wronskian. When expanding $u(\tilde{k})$ around $k$ the two expressions simplify to
\begin{eqnarray}
\beta &=& \frac{u_k' D_L u_k - u_k D_L u'_k}{W} + {\cal O}(\tilde{k}-k)^2 \\
\alpha &=& 1 + {\cal O}(\tilde{k}-k).
\end{eqnarray}
Therefore,
\begin{eqnarray}
-\frac{\beta^*}{\alpha} = \frac{u^{'*}_k D_L u^*_k - u^*_k D_L u^{'*}_k}{W^*} + {\cal O}(\tilde{k}-k)^2 \, .
\end{eqnarray}
If we now identify $u_k$ with $\gamma_k$ the Wronskian would be given by $2 i/(a M_p)^2$ and the last equation matches precisely with eq. \ref{c+}.
In the case of $Q_\zeta$ the derivation will go through in the same way with an extra detail. When Fourier transforming $D_L$ there will be a term of the form $\partial_k k$ which at a first glance makes the $-\beta^*/\alpha$ computed from the two approaches different. However, this term corresponds precisely to the change in the Jacobian which we also had to take into account in eq. \ref{optimus}.

With this result we close the equivalence between the three vertices of the triangle: charge transformation, coordinate change and Bogoliubov transformation.

\subsection{Patient Observers and Particle Production}\label{patient}

In order to measure the inflaton particle production induced by the emission of a long mode in a quasi-de Sitter phase, we imagine an Unruh type detector, a two state quantum system coupled to the inflaton in such a way that the absorption of an inflaton particle will trigger a transition in the state of the detector. Since the long scalar mode can be seen as a constant shift of the time coordinate, this will only be visible in the integrated particle production when measured according to some internal clock in the detector, which is independent of the expansion. Such a detector is not de Sitter invariant, and will detect a change in the integrated particle production measured according to the internal detector clock as the state changes from $\left|0\right> \to \left|0'\right>$.  Thus, this serves as a patient observer (defined in the introduction) in quasi de Sitter space if it is around for a sufficiently long time. An isocurvature mode during inflation is a prototype of this kind of detector \cite{Geshnizjani:2003cn}. However, if we consider the production of gravitons in pure de Sitter, the issue is a bit trickier. In this case, we can think of a detector localized inside the horizon, which is coupled to a light scalar field initially in its vacuum. The detector breaks de Sitter invariance and can record the particle excitations of the light test scalar field.

Using the expressions for the Bogoliubov coefficients defined in equations (\ref{beta}) and (\ref{alpha}) we obtain, in the dS limit,
\beq
\alpha_k =\frac{1}{2i} \left[\left(\left(\frac{k}{\tilde k}\right)^{3/2}-\left(\frac{\tilde k}{ k}\right)^{1/2}\right) \frac{1}{k\eta}-i\left(\left(\frac{k}{\tilde k}\right)^{1/2}+\left(\frac{\tilde k}{ k}\right)^{1/2}\right)\right] e^{i(\tilde k-k)\eta} ~,
\eeq
\beq
\beta_k =\frac{1}{2i} \left[\left(\left(\frac{\tilde k}{ k}\right)^{1/2}-\left(\frac{k}{\tilde k}\right)^{3/2}\right) \frac{1}{k\eta}-i\left(\left(\frac{\tilde k}{ k}\right)^{1/2}-\left(\frac{k}{\tilde k}\right)^{1/2}\right)\right] e^{i(\tilde k+k)\eta} ~.
\eeq
It is easy check that the normalization condition $|\alpha_k|^2-|\beta_k|^2=1$ is satisfied, and that on super-horizon scales one has indeed 
\beq
| u_{\tilde k}|^2 = \left(\frac{k}{\tilde k}\right)^3 | u_{k}|^2~,
\eeq
which proves that the form of $\alpha_k$ and $\beta_k$ are self consistent.

From the Bogoliubov transformation, we can compute the inflaton particle number, initially in the $\left| 0\right>$ vacuum when measured in the $\left| 0'\right>$ state,
\beq
N_k= \left<  |\beta_k|^2 \right>= \frac{1}{4} \left<\left[\left(\frac{\tilde k}{ k}\right)^3-2\left(\frac{k}{\tilde k}\right) +\left(\frac{\tilde k}{k}  \right) \right] \frac{1}{k^2\eta^2}+\left[\left(\frac{k}{\tilde k}\right)+\left(\frac{\tilde k}{ k}\right)-2\right] \right> ~,
\eeq
and so for sub-horizon modes, $-k\eta \to \infty$, we see that there is a finite piece that does not vanish
\beq
N_k \approx  \frac{\left<\delta_1^2\right>}{16}= \frac{1}{30}\left<\gamma_L^2\right>~.
\eeq
We notice that the effect for a sub-horizon observer is tiny. Even if our sub-horizon observer waits for what corresponds to the Page time of de Sitter, until the variance of the long modes that have left the horizon becomes larger than one $\left<\gamma_L^2\right> \gg 1$, there will be approximately one extra particle produced with Fourier mode $k$. However, it appears as if a sub-horizon observer after the Page time would discover that he is no more in the vacuum, but in a state with one excited particle of all sub-horizon Fourier modes. Since the one-particle state is orthogonal to the vacuum, the observer will see that the state has changed by an order one factor on all scales within the horizon \cite{1609.06318}.

Since a constant long mode does not change the time dependence of the metric, it is not expected to change the particle production rate or the periodic properties of the Green function under a complex phase shift in time. Therefore we expect that the new state, $\left|0'\right>$, will be thermal with the same temperature as $\left|0\right>$, and in fact a single pointlike Unruh type detector may not experience any change in the transition rates between energy levels in the detector due to absorption or emission of particles. One suspects that only when measuring the integrated particle production using an internal clock (such as an independently fluctuating isocurvature mode) can the difference in the particle production before and after adding the soft mode be seen. Therefore, one is led to think that such an effect might be accounted for as being due to some accumulated difference between some independent clock field and the adiabatic clock (the expansion) shifted by long modes \cite{Geshnizjani:2003cn} (see also \cite{Abramo:2001dd,Bonga:2015urq}), and if the effect is always measured in terms of the adiabatic clock (the expansion), then the effect is absent \cite{Unruh:1998ic,Abramo:2001dc, Geshnizjani:2002wp,Tsamis:2005bh, Losic:2005vg, Gasperini:2009wp, Frob:2017coq}.

\subsection{Is Patience Fundamentally Impossible?}\label{impatience}

In this subsection, we report a tentative conjecture that it may actually be impossible to build a realistic subhorizon instrument capable of acting as a patient observer.  We have performed two thought experiments on how one would go about measuring the effects of long modes, and each time we have been stymied by the requirement that we build our machine out of physically realizable matter.  This does not constitute a proof that a measurement device of sufficiently clever design cannot ever be envisioned, but it does give some indication that there may be a version of ``cosmic censorship" at play, preventing one from making these measurements.  This situation is reminiscent of \cite{DYSON:2013jra}, where it is argued that no machine capable of observing a single graviton may ever be constructed.

Our first attempt at a patient observer, as mentioned in the introduction, is a circular array of satellites very carefully bound together in the radial direction, keeping them attached under the expansion, but not preventing them from feeling shear effects.  When long modes are added, the spatial distance between the satellites changes by
\beq
ds^2 = a^2 \delta_{ij}dx^idx^j \to ds'^2  = a^2 (e^{\gamma_L})_{ij}dx^i dx^j~,  \label{squab}
\eeq
and so shear deformations become order one when $\left<\gamma_L^2\right> \sim 1$. 

However, if we compute the uncertainty in the location of the satellites due to quantum drift after a time $t$ by using $\Delta p_q = m \Delta x_q/t$, we notice from the uncertainty principle, 
\beq
\Delta x_q \geqslant \sqrt{t/m}
\eeq
where $m$ is the mass of the satellites.

On the other hand, the effect we want to measure is a shift in the position of the satellite of order
\beq
\frac{\Delta x}{x} \leqslant \sqrt{\left<\gamma^2\right>} \sim \sqrt{H^3 t/M_p^2}~.
\eeq
In order for the effect to be observable, we obviously require
\beq \label{dxq}
\frac{\Delta x}{x}  \geqslant  \frac{\Delta x_q}{x}.
\eeq
Using that the Schwarzschild radius of the satellites has to be less than the horizon, $r_s= G_N m \leqslant 1/H$, we find that (\ref{dxq}) implies
\beq
x\geqslant \frac{1}{H},
\eeq 
which implies that the detector must be larger than the horizon in order to measure the effect, which makes this setup a bad patient observer.

Note also that trying to alleviate the problem by adding $N$ satellites, as the Gaussian quantum noise in the measurement then goes down as $1/\sqrt{N}$, does not help since one still needs to require the that Schwarzschild radii of all $N$ satellites fit inside the horizon, $N r_s \leqslant 1/H$, and so the constraint above remains the same, independent of the number of satellites.

So long wavelength gravitons cannot be measured by this method.  We try again with a system of clocks, aiming to measure the effect on the expansion rate as measured by independent clocks discussed in the previous section.  Granting that the effect is present when treating the clocks and detectors classically, we check what happens when taking into account the quantum fluctuations of the detectors or clocks themselves. 

Here, the idea is that a shift by a long wavelength mode as in (\ref{squab}) would shift the time coordinate by $t\rightarrow t-\langle\g_L^2\rangle/H$. By using $\langle\g_L^2\rangle \sim \al_0 H^2/M_p^2 - \al_1 H^3 t/M_p^2  - \al_2 H^6 t^2/M_p^4 -\dots$ ($\al_i$ numerical loop factors), we can equally well absorb the effect in the expansion rate by taking  $H\to H +  \alpha_1 H^3/M_p^2+\alpha_2 H^6 t/M_p^4 +\dots \,$. One has to be careful in the physical interpretation of this kind of effect as discussed in the previous subsection\footnote{In fact it is a controversial point whether this is a physical effect or not \cite{Garriga:2007zk,Tsamis:2008zz}.}, but for the sake of the argument let us assume that this effect is present at a classical level and see if it will ever be possible to measure the effect in practice when including quantum effects on the detectors or clocks.

In order to measure a change in the expansion rate we imagine two freely falling satellites exchanging photons. The redshift of the photons, $z$, is related to the expansion rate, if the satellites are not too far from each other, by Hubble's law
\beq\label{Hubble}
z = H_0 (t_1-t_0)
\eeq
where $t_1$ is the time at which the signal is observed at the second satellite, and $t_0$ is when it was emitted at the first. If we want to measure a shift in the expansion rate of order $\Delta H_L/H \sim H^2 /M_p^2$, then we need to be able to synchronize the two clocks to a precision of $\Delta t_L/t  \sim H^2/M_p^2$.

On the other hand, if we want to look at the expansion rate in a region of size $L$, the two clocks need to be separated by the same distance. Regardless of the details of how these clocks register time, a general requirement is that if we want a clock that is accurate to a precision $\Delta t$, it must have an energy uncertainty $\Delta E\gtrsim1/(\Delta t)$. As recently described in \cite{clocks}, physical clocks must interact with each other gravitationally causing time dilation effects that will shift the time between clocks by an uncertain amount 
\beq
\Delta t_q \sim \frac{\Delta E}{M_p^2 L} t.
\eeq
This sets a fundamental limit to how precisely clocks may measure time.  To make this uncertainty the smallest, we take $\Delta E$ to be as small as possible, and $L$ as large as possible. Whatever the nature of our clock, it needs to be localized inside the horizon, giving $\Delta E > H$ and, for the same reason $L < 1/H$.  This implies that 
\beq
\Delta t_q \gtrsim \frac{H^2}{M_p^2} t=\Delta t_L~.
 \eeq
Therefore, quantum decoherence makes it impossible again to measure the shift in the expansion rate due to long modes. Therefore even if we assume that the effects of \cite{Geshnizjani:2003cn, Abramo:2001dd,Tsamis:2011ep,Bonga:2015urq, Basu:2016gyg} are there to measure when neglecting quantum fluctuations of clocks and detectors, in a full quantum treatment the effect appears unobservable.

 One might be tempted to conjecture that this could be an indication that there is some fundamental limitation to the size of infrared effects measured by patient observers in de Sitter.  This points to a tantalizing sort of quantum-geometric consistency, that even though the geometry of spacetime `fuzzes out' on such scales, it does so in such a way that it is exactly below the threshold of detectability for any local observer. Note that the analysis above is for exact de Sitter space and does not apply explicitly to inflation, as the long wavelength scalar modes are enhanced by a slow roll factor $1/\epsilon$ in that case.  It would be interesting to examine this more generally.

\section{Non-perturbative Results}\label{nonpert}

We are ultimately interested in the non-perturbative regime, where the contributions from many long modes accrue to alter the lowest order results.  The correlators
\beq
\langle \zeta_L^2\rangle=\int_{k_{min}}^{k_{max}}\frac{dk}{k}P(k),\quad\quad \langle \gamma_L^2\rangle=\int_{k_{min}}^{k_{max}}\frac{dk}{k}P_\gamma(k)\label{coxrelate}
\eeq
extend over the number of modes that can be seen by an observer.  Thus, even though we may be embedded in a very large universe, the value of $k_{max}$ we are able to take is limited by our current horizon size.  This corresponds to at the very most a few dozen e-folds, and as such places us squarely in the perturbative regime.  If we were willing to wait a very long time (and dark energy proves to eventually decay, placing us in asymptotic Minkowski space and giving us access to an unlimited number of e-folds of inflation), we would eventually reach the non-perturbative regime.  Note that since $P(k)>P_\gamma(k)$ for the entire duration of inflation, the scalar corrections will be the more relevant of the two effects.  This is fortuitous for us, as the tensor nature of the graviton corrections makes the expressions much more difficult.  As such, we will focus primarily on the scalar corrections first, and display the tensor generalization in section \ref{nonpertten}.

Before we launch into our analysis, let's make an honest assessment of how long it takes to reach the non-perturbative regime, when these quantities become large\footnote{As we will see, ``large'' means greater than $1/(1-n_s)^2$.}.  The CMB scales we have access to only yield around 8 e-folds, which is nowhere close to the amount necessary to see non-perturbative effects.  In general, the amount needed will depend on the model of inflation, as this sets the momentum dependence of the power spectrum.  For monomial models, for instance, where $V(\phi)=\lambda\phi^p$, we find that $\langle \zeta_L^2\rangle=\lambda N^{p/2+2}$, with $N$ the number of e-folds.  For a linear model, which is on the cusp of being ruled out or confirmed by tensor modes, $\lambda=10^{-3}$, and this becomes large for $N_{np}>200$.  Now, to have access to these scales, we would have to wait until scales of the size $k_{min}=e^{N_{np}}k_{max}$ enter our horizon, which would be $10^{103}$ years.  Recall that this also assumes a scenario where dark energy eventually decays away to Minkowski space, otherwise these scales would not ever reenter the horizon at all.  Models of the form $m^2\phi^2$ are even worse, taking $10^{8,000}$ years before we would see enough modes.  Plateau models of the form $V_0(1-e^{\phi/\Lambda})$ are worst of all, taking $10^{80,000}$ years.  

It can be amusing to compare these timescales to the timescales laid out in \cite{Dyson:1979zz} for the far far future evolution of the universe.  There, they find that matter behaves as a liquid on timescales of order $10^{65}$ years, with all chemical bonds tending to break apart and every object ultimately sphericalizing.  Black holes even of galactic size decay after $10^{100}$ years, and all elements decay to iron after $10^{1,500}$ years.  Ordinary clumps of matter decaying into black holes takes much, much longer, though.  Thus, even for the most optimistic models of inflation, building a deice capable of registering this effect will be a challenge.  Linear models would require some sort of repair mechanism that would counteract the tendency of the chemical bonds in the device to break, and the others would need to deal with parts of it collapsing into iron, releasing $\text{MeV}$ scale radiation.  These problems are in principle surmountable, though the authors graciously waive claims to any patent rights for such a device.

Another point of note is that dark energy is not necessarily a nuisance, when measuring this effect.  In fact, many of these timescales are actually longer than the Page time for our current cosmological expansion, $10^{130}$ years, meaning that dark energy is actually beneficial in trying to measure this effect.  However, these timescales are ridiculously impractical from a human perspective, and thus we regard the most promising avenue to ever this phenomenon to be from analog systems.  Like the recent success of (possibly) detecting phonon Hawking radiation from a mute hole \cite{mute}, if the analog of de Sitter space could be constructed in a condensed matter system, where the Planck scale is replaced with the $eV$ scale, this timeframe could be pushed down to years or decades.  We refrain from speculating what such a system would actually be at this moment.

In this section we will compute the correlators along the lines of those in (\ref{optimus}) without resorting to a loop expansion.  Our expressions for the correlators will depend on the quantities (\ref{coxrelate}) to arbitrarily high orders, and in this sense will be non-perturbative.  However, our results will not be exact, and hold only in several well defined approximations, which we detail now.

\begin{centering}
	\begin{figure*}[h]
		\centering
		\includegraphics[width=12cm]{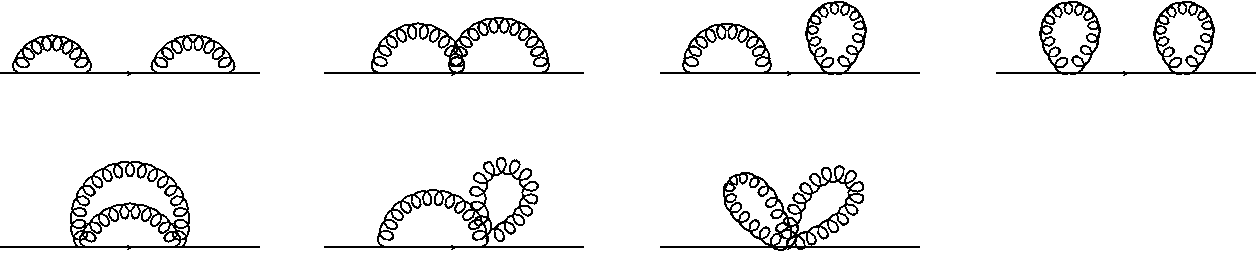}
		\caption{The two loop diagrams that are automatically included by expanding our result to second order.}
		\label{there}
	\end{figure*}
\end{centering}

\begin{centering}
	\begin{figure*}[h]
		\centering
		\includegraphics[width=12cm]{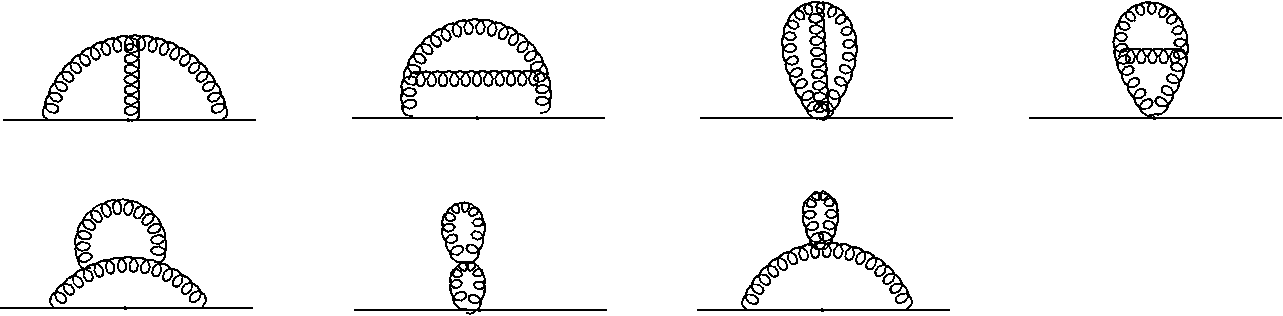}
		\caption{The two loop diagrams that are not encapsulated in our analysis.  Note that they all have self-interactions for the long modes, and so require a further insertion of the interaction Hamiltonian in the in-in formalism.}
		\label{not}
	\end{figure*}
\end{centering}

We treat the long mode state as the free field vacuum, ignoring interactions of the long modes among themselves.  Diagrammatically, this corresponds to resumming the infinite class of diagrams represented in Fig. \ref{there}, while excluding those in Fig. \ref{not}.  This amounts to treating the vacua to be those of the free fields.  So, while we include, say, fifteen loop contributions coming from the Bogoliubov coefficients, we neglect the leading corrections from self-interactions, which start at two loop order.  If desired, corrections from these higher order interactions can be added systematically to our result using the in-in formalism.

The second approximation that we make initially in order to obtain exact expressions is to treat the spectral tilt as constant, ignoring $k$ dependence.  The basic results in the non-perturbative regime will actually not be greatly affected by this approximation, and we show in section \ref{fulldist} how to easily go beyond it.

\subsection{Scalar Loops}

In this section, we show that it is possible to obtain non-perturbative results for the expectation value of correlators.  This makes it possible to probe the regime where the number of long modes becomes large, and will allow us to obtain nontrivial statements about the statistical properties of the ultra-large scale structure of the universe.  We focus here on the corrections coming from scalar modes, both because the effect is larger than that coming from tensors, and because the results are much simpler.  In the next section we return to the tensor contribution.

To illustrate our approach, we start by computing the 2 point correlator for scalar fields.  From (\ref{optimus}), the expression in the shifted coordinates is
\beq
\langle \tilde P_k\rangle=\langle e^{-2\nu \zeta_L} e^{3\zeta_L}\rangle P_k=\langle e^{(1-n_s)\zeta_L}\rangle P_k
\eeq
so that we must find the expectation value of $e^{(1-n_s)\zeta_L}$ in order to find the average power spectrum measured.  In order to do this, we write this as a path integral:
\beq
\langle e^{(1-n_s)\zeta_L}\rangle=\prod dk \int d\zeta_k \frac{e^{-\frac{\zeta_k^2}{2\sigma_k^2}}}{\sqrt{2\pi\sigma_k^2}}e^{(1-n_s)\zeta_L} \, .
\eeq
Here, we have made the two aforementioned simplifications:  that the path integral is performed in the free field vacuum, so that we can ignore interactions, and that the spectral tilt is constant over the field range under consideration.  We will comment on, and relax, each assumption in the next section, but for now we stick to this simplistic case in order to arrive at analytically solvable results.  The qualitative behavior we find extends to the more complicated cases as well.  Then, we take $\zeta_L$ to be a sum of long wavelength modes, $\zeta_L=\int_{k_{min}}^{k_{max}} dk \zeta_k$.  The path integral for momenta outside these modes trivially evaluates to 1, and the remainder are a simple Gaussian integration that yields
\beq
\langle e^{(1-n_s)\zeta_L}\rangle=e^{\frac12(1-n_s)^2\langle\zeta_L^2\rangle} \, .
\eeq
We stress that this is a non-perturbative result, valid even in the regime with a large number of modes, where the loop expansion breaks down.  Expanding the exponential in a Taylor series is equivalent to a loop expansion, and indeed it reproduces the results found in \cite{1005.1056}.  In fact, this expression was already obtained in \cite{1103.5876}, using quite different methods, and therefore serves as a nice consistency check of our methods.  The approach we take in this paper allows us to generate results like this in a streamlined fashion, which will allow us to make more sophisticated statements about observable quantities than appear in the literature.  As a first implementation of our analysis, note that this quantity is always greater than 1, indicating that the average measured power spectrum is always greater than the bare power spectrum, especially in the non-perturbative regime.  We will make more precise comments about how to interpret this in section \ref{fulldist}.  

We can compute the correction to the tensor correlator using the exact same procedure, which gives
\beq
\langle \tilde P_k^\gamma\rangle=e^{\frac12 n_t^2\langle \zeta_L^2\rangle} P^\gamma_k.
\eeq
Thus, we find the average value of the tensor to scalar ratio in the presence of a large number of long wavelength modes to be
\beq
\langle\tilde r\rangle=r\, e^{\frac12(n_t^2-(1-n_s)^2)\langle \zeta_L^2\rangle}\label{ravg} \, . 
\eeq
Though we would observe that both the tensor and scalar power increase, the ratio would either go to zero or infinity, depending on the sign of the exponent.  Using the standard slow roll consistency relations, the threshold between the two different behaviors is $r=8(1-n_s)$, which, at the best fit value for the tilt, becomes $r=.32$.  Since we can rule out tensors at this level, we conclude that, (assuming constant tilt), the brand of inflation that gave rise to our universe leads to a vanishing mean tensor to scalar ratio on extremely large scales.

\subsection{Local non-Gaussianity}
We now compute the squeezed limit of the bispectrum, to see how Maldacena's consistency relation is modified in the presence of a large number of modes.  This involves a double expansion, as the short wavelength modes are in the presence of the single soft mode of intermediate wavelength, as well as the far infrared modes.  Then 
\begin{eqnarray}
\langle \zeta_{\tilde q}\zeta_{\tilde{ k_1}}\zeta_{\tilde{k_2}}\rangle=\left\langle\left\langle\zeta_{\tilde q}e^{(1-n_s)\zeta_{\tilde q}}\right\rangle \zeta_{\tilde{ k_1}}\zeta_{\tilde{k_2}}\right\rangle=-\left\langle\frac{\partial}{\partial n_s}\left\langle e^{(1-n_s)\zeta_{\tilde q}}\right\rangle \zeta_{\tilde{ k_1}}\zeta_{\tilde{k_2}}\right\rangle\\
=(1-n_s)\left\langle\left\langle\zeta_{\tilde q}^2\right\rangle e^{\frac12(1-n_s)^2\langle\zeta_{\tilde q}^2\rangle} \zeta_{\tilde{ k_1}}\zeta_{\tilde{k_2}}\right\rangle  \, . 
\end{eqnarray}
Where we have used the expression for two point functions in a shifted background to arrive at the last line.  At this point we neglect the quantity in the exponential, as it will always be extremely small, even in the non-perturbative regime.  Then 
\beq
\langle \zeta_{\tilde q}\zeta_{\tilde{ k_1}}\zeta_{\tilde{k_2}}\rangle=(1-n_s)e^{\frac12(1-n_s+1-n_q)^2\langle\zeta_L^2\rangle}\langle\zeta_q^2\rangle\langle\zeta_{k_1}\zeta_{k_2}\rangle \, .
\eeq
So that the background modes always enhance the expected power in the three point function over the standard result.  Here we have denoted $n_q$ as the tilt at the scale of the long mode, even though in our analysis we have taken it to be constant, to emphasize the fact that both scales enter into the corrections.

The correction to the parameter $f^\text{local}_{NL}$ is
\beq
\langle\tilde{f}^\text{local}_{NL}\rangle=\frac{5}{12}\frac{\langle \zeta_{\tilde q}\zeta_{\tilde{ k_1}}\zeta_{\tilde{k_2}}\rangle}{\langle\zeta_{\tilde{q}}^2\rangle\langle\zeta_{\tilde{k_1}}\zeta_{\tilde{k_2}}\rangle}=\frac{5}{12}(1-n_s)e^{(1-n_s)(1-n_q)\langle\zeta_L^2\rangle}=f_{NL}^\text{local} e^{(1-n_s)(1-n_q)\langle\zeta_L^2\rangle}\label{fnlavg} \, ,
\eeq
in agreement with the one-loop result of \cite{1005.1056}, when expanding the exponential function to first order. We see that some terms in the exponential are not cancelled, as a consequence of its quadratic nature.  From this we infer that the long modes generically induce larger average nongaussianity, and that deep in the non-perturbative regime this quantity diverges exponentially.  This is provided that the spectrum is red everywhere:  from our expression, it appears as if it becomes blue for a period of inflation, the average nongaussianity associated with those scales will actually shrink to 0 on very large scales.  As we will discuss in the following section, this is an artefact of the constant tilt approximation.

\subsection{Tensor Loops}\label{nonpertten}

Now that we have demonstrated that our formalism can yield non-perturbative results for scalar modes, we use it to extend these results by computing the tensor contribution, using the same approximations.  Though the final expression is not quite as simple for tensors, the path integral still simplifies dramatically to an integral expression that automatically resums all graviton loops to the two point correlator.  For a single long mode this becomes a single integral of special functions, and can effortlessly be expanded to any desired order, making it a compact generating expression for all higher order loop corrections.  Including multiple long modes complicates the expression, but simplifications are still displayed.  General properties of this integral can be investigated, yielding non-perturbative results on the effect of infrared loops on the two point correlator.

To begin with, we consider the presence of a single long wavelength graviton, which serves to simplify our analysis.  Afterwards, we generalize to the case where we consider the sum of modes, so that we can study the behavior as the number of modes becomes large.

The first technical hurdle is the fact that the correlator (\ref{grimlock}), upon Taylor expansion of the exponential, involves an infinite number of distinct objects, which we labelled as $\delta_m=\gamma^m_{ij}\hat{k}_i\hat{k}_j$.  Fortunately, they are not all independent: since $\gamma$ is a $3\times3$ matrix, we can use the Cayley-Hamilton theorem, which states that every matrix obeys its own characteristic equation:
\beq
\gamma^3+\frac12[\gamma]\gamma^2+\frac12([\gamma]^2-[\gamma^2])\gamma+\det\gamma=0 \, .
\eeq
The notation $[A]=\text{trace}(A)$ is employed.  Because we are using the gauge where $\gamma$ is transverse and traceless, both the trace and determinant of $\gamma$ vanish, so this expression simplifies considerably:
\beq
\gamma^3=r^2\gamma 
\eeq
where, because we are considering only a single mode, we have $r^2=[\gamma^2]/2=\gamma_+^2+\gamma_\times^2$.  This can then be used to give a recursive relation for the different contractions, $\delta_m=r^2\delta_{m-2}$, which may be used separately for the even and odd powers of $\gamma$.  Then, the expression for the matrix exponential simplifies to 
\beq
e^{\gamma}{}_{ij}\hat k_i\hat k_j=1+\frac{\sinh r}{r}\delta_1+\frac{\cosh r-1}{r^2}\delta_2 \, . \label{hype}
\eeq
This can be simplified further by noting that $\delta_2=r^2s^2$, $s$ being the sine of the angle between momenta, and denoting $\g_+=r\cos\psi$, $\g_\times=r\sin\psi$ to yield
\beq
e^{\gamma}{}_{ij}\hat k_i\hat k_j=1+s^2\big(\cosh r-1+\sinh r\cos(\psi-2\phi)\big) \, .
\eeq
Here, $\phi$ is the angle between the momentum $k$ projected onto the plane perpendicular to some fixed unit vector $p$, the choice of which will drop out of our final expressions.

Now we must evaluate the correlator.  For a generic function of the graviton, we have
\beq
\langle f(\gamma_{ij})\rangle=\int \frac{\mathcal{D}\gamma_k}{\sqrt{2\pi\langle \gamma_L^2\rangle}}e^{-\frac12 \gamma(k)_{ij}D(k)_{ijkl}\gamma(k)_{kl}}f(\gamma_{ij}) \, .
\eeq
Again, we have treated the graviton as being in the free-field vacuum so that the correlator reduces to a Gaussian integral.

Because we are considering only a single momentum for the moment, the majority of the integrations in the path integral become trivial, and the only surviving integration reduces to
\beq
\langle f(\gamma(k)_{ij})\rangle=\int d\gamma(k)_+d\gamma(k)_\times \frac{e^{-\frac{r^2}{2\langle \gamma_L^2\rangle}}}{2\pi\langle \gamma_L^2\rangle}f(\gamma_+,\gamma_\times) \, .  \label{almost}
\eeq
We see that the kernel of integration only depends on the radial combination $r$, so that the only angular dependence comes from the quantity to be evaluated in the correlator.  For a power of the matrix exponential, both the radial and angular integrals cannot be done explicitly at the same time, but the angular one can be done, yielding a compact expression that generates the loop corrections systematically.  

Our final expression for the correlator can be written as
\beq
\left\langle\hat{\tilde{k}}^{-2\nu}\right\rangle=\int_0^\infty\frac{dr r}{2\langle \gamma_L^2\rangle} e^{-\frac{r^2}{2\langle \gamma_L^2\rangle}}R_\nu(r,s)\label{full}
\eeq
where here
\beq
R_\nu(r,s)=T_-^{-\nu}{}_2F_1\left(\frac12,\nu,1,1-\frac{T_+}{T_-}\right)+T_+^{-\nu}{}_2F_1\left(\frac12,\nu,1,1-\frac{T_-}{T_+}\right)
\eeq
and $T_\pm=1+s^2(e^{\pm r}-1)$.
Using a computer, it is trivial to expand this expression around $r=0$ to recover the loop corrections order by order.  These agree with the results obtained previously in \cite{1005.1056}.  To illustrate the power of this expression, we display the four loop correction to the power spectrum:
\ba
\left\langle\hat{\tilde{k}}^{-2\nu}\right\rangle=1+\frac{(1-n_s)(4-n_s)}{15}\langle \gamma_L^2\rangle\bigg[1+\frac{1-5n_s+n_s^2}{21}\langle \gamma_L^2\rangle\nonumber\\
+\frac{311 + 175 n_s + 590 n_s^2 - 250 n_s^3 + 25 n_s^4}{15015}\langle \gamma_L^2\rangle^2\nonumber\\
+\frac{-8781 - 14595 n_s - 3331 n_s^2 - 1875 n_s^3 + 2375 n_s^4 - 525 n_s^5 + 
 35 n_s^6}{765765}\langle \gamma_L^2\rangle^3\bigg] \, .
\ea

This is not a very useful expression, since it is doubtful we will ever be able to observe even the one loop correction, but it demonstrates that this reformulation of the problem has something new to offer.  One important thing to note, however, is that each term in this expression aside from the zeroth is proportional to $1-n_s$, indicating that in the exact de Sitter limit this expression reduces to a constant.  This will be rigorously shown in the next subsection.  What is more important is having the full expression (\ref{full}), which allows us to study its properties to gain insights into the effects of long wavelength modes on the power spectrum at a non-perturbative level.

\subsection{The de Sitter Limit}
While it is somewhat trivial to take the de Sitter limit of the tensor loop expression (\ref{full}), the tensor expression is more nontrivial.  However, since we have the full expression, it also becomes possible to address what happens to loop corrections in the de Sitter limit $n_s-1\rightarrow0$.  Though it may have been guessed that the effect will vanish, it still is comforting to verify explicitly.  To do so, it behooves us to do the angular average of (\ref{almost}) first, since the corrections only vanish after this procedure.  Taking the average, we find
\ba
\left\langle\hat{\tilde{k}}^{-2\nu}\right\rangle=\frac12\int_{-1}^1 d\mu\int_0^\infty drr\frac{e^{-\frac{r^2}{2\langle\gamma_L^2\rangle}}}{2\pi\langle\gamma_L^2\rangle}\int_0^{2\pi}d\psi\left(1+(1-\mu^2)A\right)^{-\nu}\nonumber\\
=\int_0^\infty drr\frac{e^{-\frac{r^2}{2\langle\gamma_L^2\rangle}}}{4\pi\langle\gamma_L^2\rangle}\int_0^{2\pi}d\psi\frac{{}_2F_1\left(1,\frac32-\nu,-\frac12,\frac{A}{1+A}\right)-(1+2(\nu-2)A){}_2F_1\left(1,\frac32-\nu,\frac12,\frac{A}{1+A}\right)}{(\nu-1)A(1+A)}\nn
\ea
where $A=\cosh r-1+\sinh r\cos(\psi-2\phi)$.  This expression makes it easy to see how the correlator behaves in the de Sitter limit ($\nu\rightarrow3/2$).  From the definition of hypergeometric functions, ${}_2F_1(a,0,c,x)=1$ for all $x$.  Then the integral becomes much easier to perform
\ba
\left\langle\hat{\tilde{k}}^{-2\nu}\right\rangle=\int_0^\infty drr\frac{e^{-\frac{r^2}{2\langle\gamma_L^2\rangle}}}{4\pi\langle\gamma_L^2\rangle}\int_0^{2\pi}d\psi\frac{2}{\cosh r+\sinh r\cos(\psi-2\phi)}\nonumber\\
=\int_0^\infty drr\frac{e^{-\frac{r^2}{2\langle\gamma_L^2\rangle}}}{4\pi\langle\gamma_L^2\rangle}4\pi\nonumber\\
=1  \, .
\ea

\subsection{Sum of Modes}

The results become much more complicated in the case where the background graviton is a sum of modes, but we can display some results.  This will be essential for probing the effects when the number of modes tends to infinity, as is the case for long inflation, for example.  This scenario introduces further complications to our previous reduction of the path integral, but it is possible to still make simplifications here as well.

The biggest difference is that the condition $D=\det\gamma=0$ no longer follows from the transversality condition, though $[\gamma]=0$ still holds.  This makes the recursion relation for $\delta_m$ more complicated:
\beq
\delta_m=r^2\delta_{m-2}+D\delta_{m-3} \, .
\eeq
Unfortunately, for this generalized relation the evens and odds do not split, so that the exponential (\ref{grimlock}) we are interested in does not reduce to sinhs and coshs.  Still, we can solve this equation inductively, if we set
\beq
\delta_m=A_m\delta_2+B_m\delta_1+C_m \, .
\eeq
Then the recursion relation can be written as

\begin{equation*}
\left(
\begin{array}{ccc}
A_{m+1} \\B_{m+1} \\C_{m+1}
\end{array} \right)=\mathbf{M}  \left(
\begin{array}{ccc}
A_{m} \\B_{m} \\C_{m}
\end{array} \right)=\mathbf{M}^{m+1}  \left(
\begin{array}{ccc}
0 \\0 \\1
\end{array} \right),\quad\mathbf{M} = \left(
\begin{array}{ccc}
0 & 1 & 0 \\
r^2 & 0 & 1 \\
D & 0 & 0
\end{array} \right)
\end{equation*}
and the matrix exponential can be written
\begin{equation*}
e^{\gamma}_{ij}\hat k_i\hat k_j=\sum_{m=0}^\infty\frac{1}{m!}\delta_m=(\delta_2, \delta_1, 1)e^{\mathbf{M}}\left(
\begin{array}{ccc}
0 \\0 \\1
\end{array} \right) \, .
\end{equation*}
All that remains is to exponentiate the matrix $\mathbf{M}$, and we arrive at the form
\beq
e^{\gamma}_{ij}\hat k_i\hat k_j=f_0(r,D)+f_1(r,D)\delta_1+f_2(r,D)\delta_2 
\eeq
where
\beq
f_i(r,D)=\sum_\lambda \frac{e^\lambda p_i(\lambda)}{3\lambda^2-r^2},\quad p_0=1,\quad p_1=\lambda,\quad p_2=\lambda^2-r^2
\eeq
and the sum is over the three characteristic roots of the matrix, defined by $\lambda^3-r^2\lambda+D=0$.  These functions reduce to (\ref{hype}) when the determinant is taken to be 0.

The path integral is now much harder to compute, both because the functions to be integrated are more complicated, and because with a sum over momenta the path integral does not collapse to a finite dimensional integral as readily.

\section{Full Probability Distribution}\label{fulldist}
In previous sections we demonstrated the utility of our approach by calculating several average quantities in the non-perturbative regime.  Now we turn to the fact that the average does not necessarily accurately capture what a typical observer would see.  To see this, we can compute the variance of the power spectrum, for example, by noting that
\beq
\langle \tilde P^2\rangle =\langle e^{-2(1-n_2)\zeta_L}\rangle P_0^2=e^{2(1-n_s)^2\langle\zeta_L^2\rangle}P_0^2
\eeq
so that, in the non-perturbative regime, $\sigma=e^{\frac12(1-n_s)^2\langle\zeta_L^2\rangle}\mu\gg\mu$, and the width of the distribution is very broad.  In fact, this can be extended to $\langle \tilde P^m\rangle =e^{\frac{m^2}{2}(1-n_s)^2\langle\zeta_L^2\rangle}P_k^m$.  Thus, it does not have the property of ``self-averageness'' that we encounter in situations like thermodynamics, where the distribution is so sharply peaked that we can characterize the behavior of a system by solely studying average quantities.  Fortunately, our methods allow us to go beyond computing moments, and we can recover the full probability distribution for measuring any given value of the power spectrum.  This can be found by taking the expectation value of a delta function centered a particular value of $P_k$ leading to the full distribution
\beq
p(\tilde P_k)=\frac{1}{\sqrt{2\pi(1-n_s)^2\langle\zeta_L^2\rangle}\tilde P_k}e^{-\frac{1}{2(1-n_s)^2\langle\zeta_L^2\rangle}\log^2\left(\frac{\tilde P_k}{P_k}\right)}  \, .\label{tiltdist}
\eeq
This is the log-normal distribution, which should come as no surprise, since comes from the product of exponentials of Gaussian distributions.  This produces the correct moments we computed above.  Now that we've seen the full distribution, we can elucidate some of its properties, as they pertain to the context of corrections to the power spectrum.

First, we'd like to know the limiting behaviors:  when the quantity $(1-n_s)^2\langle\zeta_L^2\rangle$ becomes small, this distribution becomes a delta function centered at the standard result.  In the opposite limit, however, it becomes a power law
\begin{equation}
p(\tilde P_k)\rightarrow \left\{
 \begin{array}{rl}
  \delta(\tilde P_k-P_k) & (1-n_s)^2\langle\zeta_L^2\rangle\rightarrow 0\\
   \frac{1}{\sqrt{2\pi(1-n_s)^2\langle\zeta_L^2\rangle}}\frac{1}{\tilde P_k} & (1-n_s)^2\langle\zeta_L^2\rangle\rightarrow \infty
 \end{array} \right.
  \, .
\end{equation}
Now it becomes clear that looking at the average quantities cannot capture the full behavior of the theory, because of the heavy tail of the distribution.  In fact, the mean of the log-normal distribution lies at an unremarkable point on the tail,  with no assurance that it will be very close to a typical observation.  

Let us now try to make the interpretation of this distribution clear:  it represents the probability that, on a given time slice set in the uniform density gauge, an observer will measure a particular value of the power spectrum.  The power spectrum is, after all, set by the microphysics of the potential as $P_k\propto V^3/V'^2$, and this does not change.  What this encapsulates, however, is where on the potential an observer would perform the measurement at that time.  The accumulation of long wavelength modes changes the scale factor locally, which in turn can be interpreted as a shift in the number of e-folds at that point.  Put another way, if one thinks of the end of inflation as being altered by the accumulation of many stochastic modes, pockets will tend to develop when inflation ends slightly earlier or later than average. 

This offers a simple way to interpret our average results (\ref{ravg}), (\ref{fnlavg}).  At a given location, the value of the physical momentum is continually jostled by long wavelength modes, resembling a Brownian process (for $\log(k)=N$).  As the number of modes becomes large, the momentum will either tend toward zero or infinity.  If the power spectrum has a constant tilt, at one of these endpoints it diverges, and thus the average of the two outcomes yields a mean result that is larger than the background value, no matter whether the spectrum is red or blue.  The same goes for $f_{NL}$, if the sign of the tilts at the two different locations are equal.  We can also understand where our result goes astray: if the spectrum is red in one portion of the spectrum and blue in another, then no matter which extreme the momentum tends toward, one of the two will go to zero, and the nongaussianity will be suppressed.  This is unphysical, however, because the actual result would keep the ratio of the two different momentum modes fixed, which when taking into account the scale dependence of the tilt would at some point give them both the same sign.

This also suggests how to go beyond the constant tilt approximation: if we have full expressions for the power spectra, we can evaluate them with the probability distribution for $k$, which we can extract from (\ref{tiltdist}) by using the change of variables $P_k=P_k^0k^{n_s-1}$:
\beq
p(\tilde k)=\frac{1}{\sqrt{2\pi\langle\zeta_L^2\rangle}\tilde k}e^{-\frac{1}{2\langle\zeta_L^2\rangle}\log^2(\tilde k/k)} \, .
\eeq
Even though we used the constant tilt expression (\ref{tiltdist}) to derive this, it is more general.  Now, if we want to transform this to a distribution for the physical observable $P_k$, we see that the prefactor is actually unchanged, as it just comes from the Jacobian of the transformation.  Thus in the non-perturbative regime the power spectrum is given by a power law distribution, with the same form as indicated before.  However, what does change is the quantity in the exponential, which is now interpreted as $k(P_k)$, which requires inverting the power spectrum (which may or may not be possible).  This complicates the condition for the onset of the non-perturbative regime, but not the scaling behavior.

\begin{centering}
	\begin{figure*}[h]
		\centering
		\includegraphics[width=16cm]{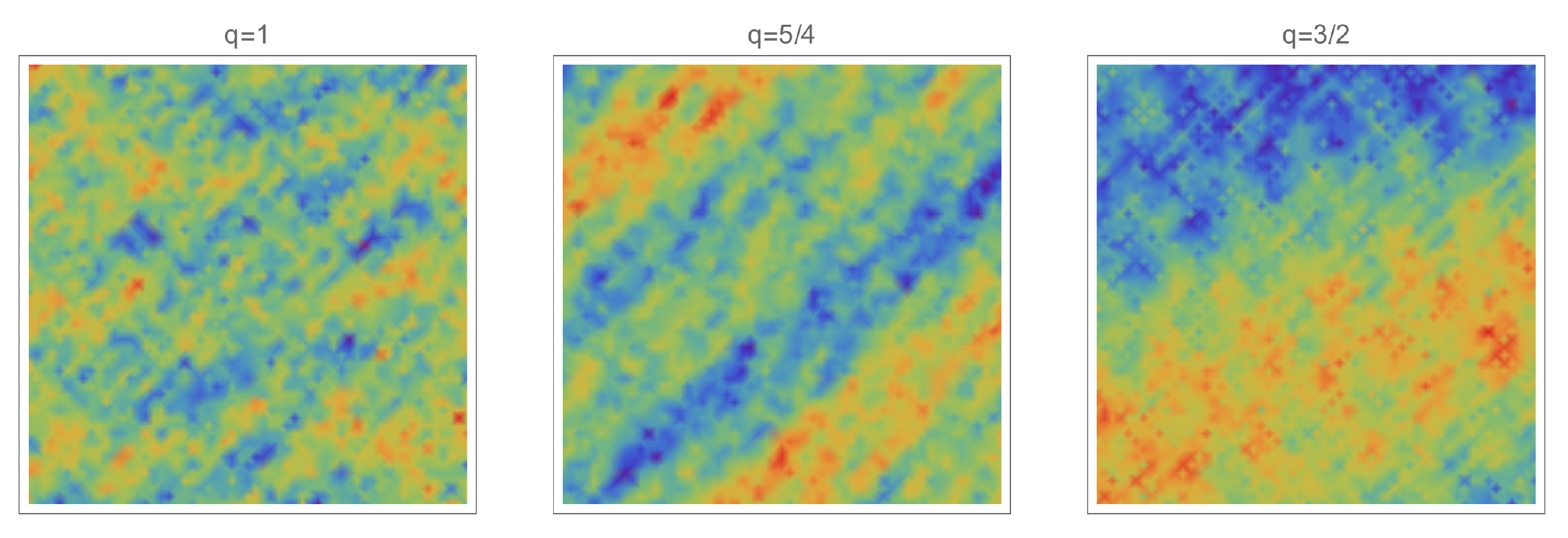}
		\caption{The ultra-large scale distribution for the number of e-folds at a given time, in increasing levels of heterogeneity.  Since the probability is simply proportional to the tilt, for weak dependence on the number of e-folds lots of little patches are intermingled, while strong dependence gives rise to massive domains.  Here each pixel represents a horizon-sized region, and the color indicates the strength of the power spectrum measured.}
		\label{distribution}
	\end{figure*}
\end{centering}

The actual scaling behavior for the power spectrum can be written simply in the non-perturbative regime, and depends on the precise model of inflation, since the tilt can be written as a function of the power spectrum.  Then
\beq
p( P_k)\rightarrow \frac{1}{\sqrt{2\pi\langle\zeta_L^2\rangle}}\frac{1}{\lvert n_s(P_k)-1\rvert P_k} = \frac{1}{\sqrt{2\pi\langle\zeta_L^2\rangle}}\bigg\lvert\frac{dN}{dP_k}\bigg\rvert \, .
\eeq
This simple expression suggests that knowledge of the $N$ dependence of the spectral tilt determines the ultra-large scale structure of the universe.  This usually comes in the form of a power law $p(P_k)\rightarrow P_k^{-q}$, and we have tabulated this dependence for several representative models in Table 1.  In Fig. \ref{distribution} we display realizations of the ultra-large scale distribution of e-folds, for different power law dependences.  These look quite qualitatively different, and indicate that an accurate determination of the functional dependence on the tilt allows us to distinguish between extremely distinct possibilities for the organization of the universe as a whole. 

\begin{table}[h]
\vskip.4cm
\begin{center}
{\renewcommand{\arraystretch}{1.5}
\renewcommand{\tabcolsep}{0.2cm}
\begin{tabular}{|c|c|c|}
\hline Model & $m$ & $q$ \\
\hline
$\lambda\phi^n$ & $\frac12$ & $\frac{4+3n}{4+2n}\in\left(1,\frac32\right)$\\ 
$V_0-\lambda\phi^n$ & $\frac{2n-2}{n-2}$ & $\frac{5n-4}{4n-4}$ \\
plateau inflation & 2 & $\frac54$\\
\hline
\end{tabular}}\label{tableone}
\end{center}
\caption{Representative values of the power law dependence for various models of inflation.}
\end{table}

Additionally, recall that we made a further approximation by omitting diagrams of the form displayed in Fig. \ref{not}.  We now argue that these do not significantly alter the form of \ref{tiltdist}, even though they are not hierarchically suppressed with respect to the diagrams we sum over.  This is an initial cause for concern, especially since the log normal distribution we have found is not one of the stable distributions guaranteed to retain its form for a suitable class of random variables.  Instead, it is known in the statistics literature \cite{tail} that correlations between the gaussian variables do deform even the tail of the log normal distribution.  These deformations can be seen to be subleading, however, by noting that corrections away from the free-field vacuum are encapsulated by the in-in formalism in the form
\beq
\langle\mathcal{O}\rangle=\langle\mathcal{O}\rangle_0-i\langle[H_{int},\mathcal{O}]\rangle_0+\dots
\eeq
for any operator $\mathcal{O}$.  The interaction Hamiltonian for gravity will contain an infinite number of terms of the form $\zeta^m\gamma^n$, and these will serve to act on the original result as derivatives, such as $\langle\zeta\mathcal{O}\rangle=P_\zeta\partial_\zeta\langle\mathcal{O}\rangle$, and so on for higher order corrections.  However, when acting on the log normal distribution, derivatives turn into terms that are suppressed by powers of $\log(P_k)/P_k$.  Thus, the direct dependence on $P_k$ cancels, and the leading contribution is logarithmic.  However, in inflation the interaction terms are suppressed by the slow roll parameters, and so these subleading results will only become important when $\log(P_k)>1/\epsilon$. In appendix \ref{appendix} we provide more arguments for neglecting such diagrams which further shows that our results are robust.

Before ending this subsection, it is important to point out that one should interpret the probability distribution that we have discussed with care. It is the probability distribution of the comoving curvature perturbation at some fixed global comoving time slice, however a local observer can only measure according to a locally defined clock and have no access to a globally defined clock. Therefore the probability distribution, which is relevant for describing an observer like us, who can only see the last $60$ e-folds of inflation, is different from the probability distribution of the curvature perturbation in the entire inflated patch. This point was discussed in more depth in \cite{1103.5876,Bartolo:2007ti,Salopek:1990re,Salopek:1990jq}, while here we will be more interested in the globally defined probability distribution, which, in the following subsection, we will use to find a criteria for eternal inflation.

\subsection{Reheating Volume}

As an application of our probability distribution, we use it to calculate the expected reheating volume for some simple inflationary scenarios.  This is used as a diagnostic for eternal inflation, as if this quantity diverges in the late time limit we can tell that inflation never ends.  As in \cite{0802.1067}, the reheating volume can be expressed as
\beq
\langle V\rangle=V_0\int d\Delta N \,p(\Delta N) \, .
\eeq
Here, $V_0$ is some arbitrary initial volume, which we have pulled outside the integral using statistical homogeneity.  Additionally, we have neglected the fact that once a portion of the universe reheats, the Brownian jitter of long wavelengths ceases, precluding that portion from spontaneously undergoing inflation again, but this was shown in \cite{0802.1067} to give a negligible contribution to the overall evolution.  Then we can use our expression \ref{tiltdist}: when expressed in terms of number of efolds, it simply becomes a Gaussian distribution
\beq
p(\Delta N)=\frac{1}{\sqrt{2\pi\langle\zeta_L^2\rangle}}e^{-\frac{\Delta N^2}{2\langle\zeta_L^2\rangle}} \, .
\eeq
To specify a model of inflation, we need the dependence of the power spectrum on time, so that we can use \ref{coxrelate} to find the variance of the long modes.  If we take the ansatz $P_k=P_K^0(N/N_0)^m=P_k^0(1+\Delta N/N_0)^m$, then $\langle \zeta_L^2\rangle=N_0P_k^0/(m+1)(1+\Delta N/N_0)^{m+1}$.  This can then be used to find the average reheating volume:
\beq
\langle V\rangle=V_0\int d\Delta N A(\Delta N)e^{B(\Delta N)},\quad B(\Delta N)=3\Delta N-\frac{m+1}{2N_0}\frac{\Delta N^2}{(1+\Delta N/N_0)^{m+1}} \, .
\eeq
Here $A(\Delta N)$ is some sub-exponential prefactor that does not determine the convergence properties of the integral.  What does determine the convergence is the asymptotic behavior of the quantity inside the exponential: if $B(\Delta N)$ asymptotes to $-\infty$ as $\Delta N\rightarrow + \infty$, the integral converges.  Otherwise, it does not.  The conditions for this are rather simple: since the quantity is 
\beq
B(\Delta N)\rightarrow3\Delta N-\frac{m+1}{2}N_0^m \Delta N^{1-m} 
\eeq
the integral converges for $m<0$, and diverges for $m>0$.

The case $m=0$ is more subtle, despite it being the most discussed in the literature, as it corresponds to a perfectly flat power spectrum.  Then, if $P_k=P_k^0$, $\langle\zeta_L^2\rangle=P_k^0\Delta N$, and
\beq
B(\Delta N)=\left(3-\frac{1}{2P_k}\right)\Delta N \, .
\eeq
In this case both terms are linear, and the convergence depends on the coefficient:  if $P_k<1/6$, inflation proceeds classically, while if $P_k>1/6$ inflation is in the eternal regime, and quantum fluctuations dominate the evolution.

This standard analysis has been used to diagnose many inflationary scenarios, for instance in \cite{Barenboim:2016mmw}, where they concluded that hilltop inflation is capable of yielding eternal inflation.  What this does not take into account is that the strength of the fluctuations can vary with location on the inflationary potential, and thus time.  This makes an analysis taking the time dependence of the power spectrum more realistic.  For instance, in the parameterization we have chosen, $m=1/(3-2q)$ from Table 1, so that the condition for $m$ to be negative is $q>3/2$.  For hilltop inflation, with  with $V(\phi)=V_0-\lambda\phi^n$, this occurs except for the range $1<n<2$, confirming the claims of \cite{Barenboim:2016mmw} that hilltop inflation can generically be eternal.  Also, plateau inflation can be seen as a kind of limiting case of hilltop inflation in this regard, giving the same distribution as the $n\rightarrow\infty$ limit.  Monomial models are always eternal, independent of the power of the field that appears in the potential.

One other feature of this expression to note is that the models which yield eternal inflation do so independently of the starting conditions.  This may seem to contradict standard intuition that they only achieve this behavior if the initial field value is above some critical threshold.  However, the quantity we are calculating is the reheating volume averaged over many realizations of the dynamics.  There will always be a (potentially very small) probability of fluctuations pushing the inflaton field further up the potential than its starting point, all the way to the eternal regime.  Once this happens, the volume in this realization is infinite, and because the probability does not vanish, the quantity will be dominated by this realization.  Other diagnostics must be used if one is interested in the relative likelihood of eternal inflation for a given set of initial conditions, but this quantity gives us information of whether eternal inflation is possible at all for a given potential.

\section{Conclusions}

In this work we have shown that a coordinate transformation, a transformation generated by the charge associated with the asymptotic symmetries of de-Sitter and a Bogoliubov transformation are three equivalent ways of capturing the infrared effects on the local vacua in de-Sitter. 

Even though the perturbative description of inflation breaks down at the Page time, we have shown that making use of the equivalence above it is possible to obtain non-perturbative results that accurately reflect the dynamics of the system.  This is done by resumming an infinite class of diagrams, which results in explicit expressions for the case of long wavelength scalars, and  highly simplified expressions for the more complicated case of tensors.  We used this to explicitly compute the physical quantity corresponding to the probability that a given value of the power spectrum is measured on a flat time slice.  In principle, this method may be used to calculate other observables, such as the fractal dimension of the reheating surface, as in \cite{Aryal:1987vn,Vilenkin:1992uf,gr-qc/0111048}.

We have also used our calculations to investigate if a physical effect can in principle be observed by a local observer confined to a single Hubble patch\footnote{Here we are not considering the possibility of the de Sitter expansion ending and modes re-entering the horizon, as discussed in \cite{Giddings:2011zd}.} , if they are willing to patiently tend to their apparatus for a long enough time. We found that the answer is non-trivial when including quantum fluctuations of the detector, and appears to obstruct the measurement of an effect even for a patient observer in de Sitter space. It would be interesting to generalize this discussion, to test if there is a new type of UV/IR cosmic censorship for patient observers in de Sitter, as we are tempted to conjecture. 

\subsection*{Aknowledgements}

We would like to thank Paolo Creminelli, Jaume Garriga, Steve Giddings, Atsushi Higuchi, Nemanja Kaloper, Alexandros Kehagias, Antonio Riotto and Sergey Sibiryakov for interesting discussions or comments. RZF is supported by ERC Starting Grant HoloLHC-306605. MSS is supported by Villum Fonden grant 13384. CP3-Origins is partially funded by the Danish National Research Foundation, grant number DNRF90.

\appendix 

\section{Corrections from Soft Mode Self-Interactions \label{appendix}}

In section \ref{nonpert} when deriving non-perturbative results we summed over all interactions between hard and soft modes. One could worry that other corrections like the ones coming from soft mode self-interactions in fig. \ref{not} could spoil the results obtained. In this appendix we argue why these diagrams are negligible compared to the diagrams in fig. \ref{there}.

Consider, for instance, the 7th diagram in fig. \ref{not} which has 2 vertices. For ease of notation, let us only consider scalar interactions. Thus, one of the vertices involves a soft-hard interaction of the form
\begin{eqnarray}
\int dt \p \zeta_k \p \zeta_q \p \zeta_q (t) , \qquad \text{Vertex 1} \equiv V_1
\end{eqnarray}
where $k \rightarrow 0$ is the momentum of the soft mode and $\p$ is any type of derivative, and the other involves only soft modes
\begin{eqnarray}
\int dt'  \p \zeta_k \p \zeta_k \p \zeta_k (t') , \qquad \text{Vertex 2} \equiv V_2.
\end{eqnarray}
We would like to compare a diagram with these 2 vertices with a diagram involving two vertices of the type 1.
As is well known, when a given mode exits the horizon it becomes classical reflecting the fact that the imaginary part of the mode function becomes suppressed compared to the real part.
In the in-in formalism, due to its commutator structure,  we have to pick at least one imaginary part of one mode function in each vertex. This means that one of the mode functions in the second vertex has to be imaginary. Thus, if all legs in the second vertex are soft in the whole time integration then the ratio between the two referred diagrams would be
\begin{eqnarray}
\frac{V_2}{V_1} \simeq \frac{k\, \text{Im}[\zeta_k]}{q \,\text{Im}[\zeta_q]}  \lesssim \left(\frac k q \right)^4 
\end{eqnarray}
where we assumed that the support of the time integration comes from the horizon crossing time of the shortest mode.

Although this argument shows that loops of soft modes are suppressed compared to loops involving soft and hard modes the computation at 2-loop is not that simple. At 1-loop it was shown in \cite{1005.1056} that the IR divergences cancel out in de-Sitter and single-field inflation once we sum over all diagrams. At 2-loop there are mixed IR-UV divergences which need to be treated carefully. However, such a computation goes beyond the scope of this paper.

\bibliographystyle{JHEP}
\bibliography{bogbib.bib}

\end{document}